\documentclass[12pt]{article}

\begin{document}

\title{Value of color charges and structure of gauge bosons}
\author{Amjad Hussain Shah Gilani \\
National Center for Physics, Quaid-i-Azam University,\\
Islamabad 45320, Pakistan}
\date{}
\maketitle

\begin{abstract}
The values of color and anticolor charges are proposed. The structure of
gluons is predicted relative to their color and anticolor charges. It is
shown that the gauge bosons of lower order theories can be used as it is for
higher order theories. Various mass relations between the gauge bosons are
also given.
\end{abstract}

\section{Introduction}

Since the birth of Quantum Chromodynamics (QCD), no attention was paid to
find out the numerical value of the color charges. The color combinations
were given in the literature with the help of SU(3) on the anology of quark
combinations and there was not clear definition for color and anticolor. It
was observed that the gluons cannot carry a combination of a color and
anticolor charge, because this combination violate the group property \cite
{gilani1}. Gilani \cite{gilani1} pointed out that gluons will carry either
color charges or anticolor charges. He claimed, with the help of set theory,
that there will be six colored gluons, one color singlet gluon and one
massless neutral gluon. He also claimed that seven gluons will be massive
and only one massless gluon. Relations between color charges and anticolor
charges are also defined in Ref. \cite{gilani1}.

The paper is organized as follows: In sec. \ref{value}, the values of the
color charges are predicted and obtained the values of respective anticolor
charges from color charges as defined in Ref. \cite{gilani1}. Charged gluons
structure is predicted in Sec. \ref{structureofgluons}. Charged gluons
masses are predicted in Sec. \ref{masses}. Pictorial representation of the
structure of electroweak gauge bosons and gluons is presented in Sec. \ref
{pictorial}. Position of massive gluon is explained in Sec. \ref{position}.
Section \ref{higgs} is devoted to Higgs. Relation between electroweak gauge
bosons and QCD gauge bosons (gluons) is discussed in Sec. \ref{relation}.
Expected Higgs mass is predicted in Sec. \ref{expected} by using the
existing value of $W$-boson mass $\left( m_W\right) $. Quark color charges
and some decays of mesons and baryons are presented in Sec. \ref
{Quarkcolorcharges}. Finally, the results are summarized in Sec. \ref
{conclusions}.

\section{Can we give values to color charges ? \label{value}}

The electroweak charges are two i.e. $+1$, $-1$. The question is: Is there
any mathematical relation which has such a solution that we obtain a
solution set as $\left\{ +1,-1\right\} $ ? The equation is 
\begin{eqnarray}
x^2 &=&1  \label{x2-1} \\
x &=&\left\{ +1,-1\right\}  \label{x2-2}
\end{eqnarray}
This shows that the electroweak charges are the result of square roots of
unity. In QCD, there are three charges and if we take the cube roots of
unity, we obtain 
\begin{eqnarray}
x^3 &=&1  \label{x3-1} \\
x &=&\left\{ +1,-\frac 12\pm i\frac{\sqrt{3}}2\right\}  \label{x3-2}
\end{eqnarray}
that is, we can predict here that the color charges are equivalent to cube
roots of unity and we define 
\begin{eqnarray}
r &\equiv &+1  \label{c1} \\
g &\equiv &-\frac 12+i\frac{\sqrt{3}}2  \label{c2} \\
b &\equiv &-\frac 12-i\frac{\sqrt{3}}2  \label{c3}
\end{eqnarray}
On the other hand, the anticolor charges can be obtain from the color
charges and we obtain 
\begin{eqnarray}
\bar{b} &\equiv &r+g=+\frac 12+i\frac{\sqrt{3}}2  \label{c4} \\
\bar{g} &\equiv &b+r=+\frac 12-i\frac{\sqrt{3}}2  \label{c5} \\
\bar{r} &\equiv &g+b=-1  \label{c6}
\end{eqnarray}
The sum of all the color charges is equal to zero and the same is true for
anticolor charges. 
\begin{eqnarray*}
r+g+b &=&0 \\
\bar{r}+\bar{g}+\bar{b} &=&0
\end{eqnarray*}

The color and anticolor charges (Eqs. (\ref{c1}--\ref{c6})) are ploted in
Fig. 1. All the distances $oi$ ($i=r,g,b,\bar{b},\bar{g},\bar{r}$), $r\bar{b}
$, $\bar{b}g$, $g\bar{r}$, $\bar{r}b$, $b\bar{g}$ and $\bar{g}r$ are equal
to unity.

\section{Structure of charged gluons \label{structureofgluons}}

In electroweak theory, the charged vector bosons $W^{+}$ and $W^{-}$ carry
only $+1$ and $-1$ charge respectively. In QCD, as in electroweak theory,
color charged gluons carry color charges contrary to color-anticolor charge,
this was pointed out in a recent study \cite{gilani1}. In electroweak, there
are only two charges and hence we have only two charged vector gauge bosons
i.e. $W^{+}$, $W^{-}$. In QCD, there are three color charges and their
corresponding three anticolor charges as defined in Eqs. (\ref{c1}--\ref{c6}%
). So, the color and anticolor charges gave us three colored and three
anticolored vector gauge bosons (i.e. gluons). Let us suppose that a gluon $%
g_{+}$ be associated to positive charge and a gluon $g_{-}$ be associated to
negative charge in QCD. So, we define the various color and anticolor
charged gluons with respect to their charges as 
\begin{eqnarray}
g_r &\equiv &+g_{+}  \label{g1} \\
g_g &\equiv &\left( -\frac 12+i\frac{\sqrt{3}}2\right) g_{-}  \label{g2} \\
g_b &\equiv &\left( -\frac 12-i\frac{\sqrt{3}}2\right) g_{-}  \label{g3} \\
g_{\bar{b}} &\equiv &\left( +\frac 12+i\frac{\sqrt{3}}2\right) g_{+}
\label{g4} \\
g_{\bar{g}} &\equiv &\left( +\frac 12-i\frac{\sqrt{3}}2\right) g_{+}
\label{g5} \\
g_{\bar{r}} &\equiv &-g_{-}  \label{g6}
\end{eqnarray}
We give the positive or negative charge to gluons only seeing the sign of
the real part, see for example Eqs. (\ref{g2}--\ref{g5}). If you are not
convinced at this stage why we give the same charge to the hybrid gluons
[Eqs. (\ref{g2}--\ref{g5})] as the sign their real part, you will be
convinced after doing exercise of decay processes in Sec. \ref
{Quarkcolorcharges}.

\section{Masses of charged gluons \label{masses}}

As in electroweak theory, the masses of charged vector gauge bosons are
equal. On same analogy we suppose that $m_{g_{+}}=m_{g_{-}}=m_g$. Therefore, 
\begin{eqnarray}
m_{g_r} &=&\left| +m_{g_{+}}\right| =m_g  \label{masses1} \\
m_{g_g} &=&\left| \left( -\frac 12+i\frac{\sqrt{3}}2\right) m_{g_{-}}\right|
=m_g  \label{masses2} \\
m_{g_b} &=&\left| \left( -\frac 12-i\frac{\sqrt{3}}2\right) m_{g_{-}}\right|
=m_g  \label{masses3} \\
m_{g_{\bar{b}}} &=&\left| \left( +\frac 12+i\frac{\sqrt{3}}2\right)
m_{g_{+}}\right| =m_g  \label{masses4} \\
m_{g_{\bar{g}}} &=&\left| \left( +\frac 12-i\frac{\sqrt{3}}2\right)
m_{g_{+}}\right| =m_g  \label{masses5} \\
m_{g_{\bar{r}}} &=&\left| -m_{g_{-}}\right| =m_g  \label{masses6}
\end{eqnarray}
Equations (\ref{masses1}--\ref{masses6}) give the masses of gluons and we
have found that the masses of color charged gluons and anticolor charged
gluons are equal, i.e. $m_{g_r}=m_{g_g}=m_{g_b}=m_{g_{\bar{b}}}=m_{g_{\bar{g}%
}}=m_{g_{\bar{r}}}=m_g$.

\section{Pictorial representation of electroweak gauge bosons and gluons 
\label{pictorial}}

The electroweak gauge bosons are $\gamma $, $W^{+}$, $W^{-}$ and $Z^0$.
Among these four gauge bosons, only photon $\left( \gamma \right) $ is
massless while the remaining three are massive. The gauge bosons are
predicted only by set theory but not by special unitary groups i.e. SU(2) or
SU(3) \cite{gilani1}. The electroweak charges can be described by Eq. (\ref
{x2-1}) and we obtain a set of its roots as given in Eq. (\ref{x2-2}). A set
has proper and improper subsets. In this case (\ref{x2-2}), we have two
component set, so we have two proper subsets and two improper subsets, i.e., 
\[
\left\{ +1,-1\right\} \Rightarrow \left\{ \left\{ \,\,\right\} ,\left\{
+1\right\} ,\left\{ -1\right\} ,\left\{ +1,-1\right\} \right\} 
\]
The subsets $\left\{ +1\right\} $, $\left\{ -1\right\} $ are proper subsets
of the set of charges while $\left\{ \,\,\right\} $, $\left\{ +1,-1\right\} $
are improper subsets. If we plot all these, then we fix the empty subset at
the origin and $\left\{ +1\right\} $ on the x-axis at $+1$ position and $%
\left\{ -1\right\} $ at the $-1$ position while the $\left\{ +1,-1\right\} $
gets the position above the empty subset along the z-axis. We have shown the
electroweak gauge bosons in Fig. \ref{electroweakbosons}. How the $Z^0$
vector gauge boson gets the position above the photon $\left( \gamma \right) 
$, we will explain it in one of the next sections.

It was claimed that the gluons does not obey the SU(3) but can be predicted
by set theory \cite{gilani1}. The value of the color charges is explained in
section \ref{value} and ploted in Fig. \ref{figvalue}. We have sketched a
pattern of the gluons with respect to their charges as shown in Fig. \ref
{figgluons}.

\section{Position of the massive neutral gluon \label{position}}

The massive color-singlet (neutral) gluon $G_0$ is placed over the massless
gluon as shown in Fig. \ref{figgluons} just like the neutral vector boson $%
Z^0$ is placed over the photon $\left( \gamma \right) $, see Fig. \ref
{electroweakbosons}. The question is, how ? and why ?

To answer this question, let us join the color points by straight lines
which results in an equilateral triangle $\Delta rgb$ having the length of
each side $\sqrt{3}$. Similarly, we can draw anticolor triangle $\Delta \bar{%
b}\bar{g}\bar{r}$ by joining anticolor points. Anticolor triangle $\Delta 
\bar{b}\bar{g}\bar{r}$ is also an equilateral triangle having length of each
side of the triangle $\sqrt{3}$ as shown in Fig. \ref{cactriangle}. Take the
apex $r$ of triangle $\Delta rgb$ and $\bar{r}$ of the other triangle $%
\Delta \bar{b}\bar{g}\bar{r}$, and join them togather, which meet on the
z-axis as shown in Fig. \ref{trianglerrbar}. The points $r$ and $\bar{r}$
meet on the z-axis at the point $z$. On the same grounds, if we take apexes $%
g$ and $\bar{g}$ of the respective triangles, which meet again at the same
point $z$ on the z-axis, similar the case will be for the apexes $b$ and $%
\bar{b}$. Now the question is, what is the value of $z=?$ Let us any
triangle $\Delta ozr$, the side $\left| or\right| =1$, $\left| rz\right| =%
\sqrt{3}$ and $\left| oz\right| =?$. The triangle $\Delta ozr$ is a right
triangle. Applying Pathagora's theorem, $\left| oz\right| ^2=\left|
rz\right| ^2-\left| or\right| ^2=3-1=2$ i.e. $z=\left| oz\right| =\sqrt{2}$.
This shows that the point $z$ is $\sqrt{2}$ units above the origin along the
z-axis.

\section{Higgs: Is there any ? \label{higgs}}

Following the discussion given in Sec. \ref{position}, in triangle $\Delta
ozr$, the side $\left| or\right| $ gives the size of the charged gluon i.e. $%
W^r\left( g_r\right) $ [see Eq. (\ref{g1})] and the side $\left| oz\right| $
gives the size of the color singlet gluon $\left( G_0\right) $. The
remaining third side of the triangle $\left| zr\right| $ will give the size
of Higgs. We get six Higgs of equal size. Among these Higgs, three carry
color charge and three carry anticolor charge. The sides of the triangle $%
\Delta ozr$ have certain ratio between each other, i.e. 
\[
\left| or\right| :\left| oz\right| :\left| rz\right| =1:\sqrt{2}:\sqrt{3} 
\]
From the above ratio between the side of triangle, we can predict the masses
of the color singlet gluon $\left( G_0\right) $ and the Higgs $\left(
H_i\right) $ as 
\begin{eqnarray}
m_{G_0} &=&\sqrt{2}m_g=\sqrt{2}m_W,  \label{neutralgluonmass} \\
m_{H_i} &=&\sqrt{3}m_g=\sqrt{3}m_W,  \label{coloredhiggs}
\end{eqnarray}
respectively. 

\section{Relation between electroweak and QCD gauge bosons \label{relation}}

Is there any relation between electroweak and QCD theories ? This question
raised by many scientists who wrote articles entitled: `Theory of Every
Thing'. Keeping in mind the above discussion, we conclude that QCD is
nothing but a higher order electroweak theory. Electroweak theory is second
order theory and QCD is third order theory because the structure of their
charges obtained from Eqs. (\ref{x2-1}) and (\ref{x3-1}) respectively. The
position of $r=+1$ and $\bar{r}=-1$ charges is exactly same as the
respective charges in electroweak theory. Due to this similarity, we suppose
that 
\begin{equation}
g_{+}=W_{+},\,\,\,\,\,\,\,\,\,g_{-}=W_{-}
\end{equation}
Therefore, we can modify the Eqs. (\ref{g1}--\ref{g6}) as 
\begin{eqnarray}
W^r &=&g_r\equiv +W_{+},  \label{g1a} \\
W^g &=&g_g\equiv \left( -\frac 12+i\frac{\sqrt{3}}2\right) W_{-},
\label{g2a} \\
W^b &=&g_b\equiv \left( -\frac 12-i\frac{\sqrt{3}}2\right) W_{-},
\label{g3a} \\
W^{\bar{b}} &=&g_{\bar{b}}\equiv \left( +\frac 12+i\frac{\sqrt{3}}2\right)
W_{+},  \label{g4a} \\
W^{\bar{g}} &=&g_{\bar{g}}\equiv \left( +\frac 12-i\frac{\sqrt{3}}2\right)
W_{+},  \label{g5a} \\
W^{\bar{r}} &=&g_{\bar{r}}\equiv -W_{-},  \label{g6a}
\end{eqnarray}
and their masses 
\begin{equation}
m_{g_r}=m_{g_g}=m_{g_b}=m_{g_{\bar{b}}}=m_{g_{\bar{g}}}=m_{g_{\bar{r}}}=m_W.
\label{mga}
\end{equation}
Also the Higgs masses are 
\begin{equation}
m_{H_i}=\sqrt{3}m_W,\,\,\,\,i=r,g,b,\bar{b},\bar{g},\bar{r}
\label{coloredhiggsa}
\end{equation}
and color singlet gluon $G_0$ mass is 
\begin{equation}
m_{G_0}=m_{Z_0}=\sqrt{2}m_W.  \label{colorsinglet}
\end{equation}

If we take away the blocks of $g_{\bar{g}}$, $g_g$, $g_b$ and $g_{\bar{b}}$
from Fig. \ref{figgluons}, we are left with Fig. \ref{electroweakbosons}.
Again if we take away the blocks of $W^{+}$ and $W^{-}$, we are left with
massless photon $\left( \gamma \right) $ and $Z^0$, which serves as the
gauge bosons for theory of gravity. This means that the photon will serve
the purpose of massless graviton ${\cal g}$ and $Z^0$ plays the role of
massive graviton $\mathcal{G}$ as predicted first time by Gilani in his
recent article \cite{gilani1}. Finally, take away the massive graviton $%
\left( Z^0\right) $, we are left with massless photon at the origin which
serve the purpose of Casimirion \cite{gilani1} as a gauge boson of Casimir
force.

\section{Expected Higgs mass \label{expected}}

Consider triangle $\Delta ozr$ (see Fig. \ref{trianglerrbar}) 
\begin{eqnarray}
\left| rz\right| ^2 &=&\left| or\right| ^2+\left| oz\right| ^2  \nonumber \\
\left| rz\right| &=&\sqrt{\left| or\right| ^2+\left| oz\right| ^2}  \nonumber
\\
m_H &=&\sqrt{m_W^2+m_Z^2}  \nonumber \\
&=&121.5855 {\,\, GeV}  \label{hnv1}
\end{eqnarray}
where $m_W=80.423$ GeV and $m_Z=91.1876$ GeV \cite{PDG2002}. Whereas, from
Eq. (\ref{coloredhiggsa}) 
\begin{eqnarray}
m_H &=&\sqrt{3}m_W  \nonumber \\
&=&139.2967{\,\, GeV}  \label{hnv2}
\end{eqnarray}
From Eqs. (\ref{hnv1}) and (\ref{hnv2}), the two values for Higgs masses
does not match. Equation (\ref{hnv1}) obtains the value of Higgs mass from
the experimental values of $m_W$ and $m_Z$, while Eq. (\ref{hnv2}) gives the
value of Higgs mass using only experimental value of $W$-mass. This scheme
gives the relation between the $m_W$ and $m_Z$ [see Eq. (\ref{colorsinglet}%
)]. This Eq. (\ref{colorsinglet}) does not match the experimental value of $%
m_Z$, if we take the $W$-boson mass as standard.

\section{Quark color charges \label{Quarkcolorcharges}}

We will not consider the quark fractional charges, like $+\frac 23e$ for
up-type quarks and $-\frac 13e$ for down-type quarks. We cannot take at the
same time two type of charges i.e. fractional charges and color charges upon
the quarks. We recommend to give color charges to up-type quarks (i.e. $u$, $%
c$, $t$) and anti-color charges to down-type quarks (i.e. $d$, $s$, $b$).
Let us see when a $B$ meson decay into a $\rho $-meson 
\begin{eqnarray}
\bar{B}^0\left( b^{\bar{g}}\bar{d}^g\right) &\rightarrow &W^r\rho ^{\bar{r}%
}\left( u^b\bar{d}^g\right) ,  \label{md1} \\
&\rightarrow &W^b\rho ^{\bar{b}}\left( u^r\bar{d}^g\right) ,  \label{md2}
\end{eqnarray}
In the above decay, a $b^{\bar{g}}$ quark decay into $u^b\left( u^r\right) $
and $W^r\left( W^b\right) $, and $W^r\equiv W^{+}$ further decay into $%
l^{+}\nu _l$. In the above decay we consider only green color and anticolor
combination. Let us see if we take the other combination of quark colors for
the decay of $B^0$-meson. 
\begin{eqnarray}
\bar{B}^0\left( b^{\bar{b}}\bar{d}^b\right) &\rightarrow &W^r\rho ^{\bar{r}%
}\left( u^g\bar{d}^b\right) ,  \label{md3} \\
&\rightarrow &W^g\rho ^{\bar{g}}\left( u^r\bar{d}^b\right) ,  \label{md4} \\
\bar{B}^0\left( b^{\bar{r}}\bar{d}^r\right) &\rightarrow &W^b\rho ^{\bar{b}%
}\left( u^g\bar{d}^r\right) ,  \label{md5} \\
&\rightarrow &W^g\rho ^{\bar{g}}\left( u^b\bar{d}^r\right) ,  \label{md6}
\end{eqnarray}
where 
\begin{eqnarray}
W^b &=&\left( -\frac 12-i\frac{\sqrt{3}}2\right) W^{-}, \\
W^g &=&\left( -\frac 12+i\frac{\sqrt{3}}2\right) W^{-},
\end{eqnarray}
and 
\begin{eqnarray}
\rho ^{\bar{g}} &=&\left( \frac 12-i\frac{\sqrt{3}}2\right) \rho ^{+},
\label{rhogbar} \\
\rho ^{\bar{b}} &=&\left( \frac 12+i\frac{\sqrt{3}}2\right) \rho ^{+}.
\label{rhobbar}
\end{eqnarray}
This shows that by solving the decays like $\bar{B}^0\left( b^{\bar{g}}\bar{d%
}^g\right) \rightarrow W^r\rho ^{\bar{r}}\left( u^b\bar{d}^g\right) $ and $%
B^0\left( \bar{b}^gd^{\bar{g}}\right) \rightarrow W^{\bar{r}}\rho ^r\left( 
\bar{u}^{\bar{b}}d^{\bar{g}}\right) $, we can solve all the remaining decays
easily. We cannot ignore the possibility that the up-type quarks serve as
quarks and their respective down-type quarks serve as anti-quarks but at the
moment we are not sure. If such a possibility exists then the quarks will be
reduced to three (i.e. $u$, $c$, $t$) and their anti-quarks (i.e. $\bar{u}%
\equiv d$, $\bar{c}\equiv s$, $\bar{t}\equiv b$), then the life will become
too much simple. We will now focus on the hybrid states like given in Eqs. (%
\ref{rhogbar}, \ref{rhobbar}), if such states are virtual and further decay.
Then 
\[
\rho ^{\bar{b}}\left( u^r\bar{d}^g\right) \rightarrow W^{\bar{b}}\rho
^0\left( d^{\bar{g}}\bar{d}^g\right) . 
\]
Following $B^0$ decay given in Eq. (\ref{md2}) 
\begin{eqnarray}
\bar{B}^0\left( b^{\bar{g}}\bar{d}^g\right) &\rightarrow &W^b\rho ^{\bar{b}%
}\left( u^r\bar{d}^g\right)  \nonumber \\
&\rightarrow &W^bW^{\bar{b}}\rho ^0\left( d^{\bar{g}}\bar{d}^g\right) 
\nonumber \\
&\rightarrow &Z^0\rho ^0\left( d^{\bar{g}}\bar{d}^g\right)
\end{eqnarray}
So, in the above decay 
\begin{equation}
b^{\bar{g}}\rightarrow W^bu^r\rightarrow W^bW^{\bar{b}}d^{\bar{g}%
}\rightarrow Z^0d^{\bar{g}}.
\end{equation}

For three quark baryon states, we will write the combinations as 
\begin{eqnarray}
\left( \frac{u^bd^{\bar{b}}+u^gd^{\bar{g}}}{\sqrt{2}}\right) d^{\bar{r}%
}=\left( udd\right) ^{-},\,\,u^r\left( \frac{u^bd^{\bar{b}}+u^gd^{\bar{g}}}{%
\sqrt{2}}\right) =\left( uud\right) ^{+},  \label{3quarkstate1}
\end{eqnarray}
In the above examples, combination of type $u^rd^{\bar{r}}d^{\bar{r}}$ or $%
u^ru^rd^{\bar{r}}$ does not exist because of repeated anticolor or color
index, repectively. We are just giving here simple examples. The rest of the
states we can make like 
\begin{equation}
\left( \frac{u^gd^{\bar{b}}+u^bd^{\bar{g}}}{\sqrt{2}}\right) d^{\bar{r}%
}=\left( udd\right) ^{-},  \label{3quarkstate2}
\end{equation}
etc. If $\Lambda _b^0\rightarrow W^{+}\Lambda ^{-}$, then 
\begin{equation}
\Lambda _b^0\left( b^{\bar{g}}d^{\bar{b}}d^{\bar{r}}\right) \rightarrow
W^r\Lambda ^{\bar{r}}\left( u^bd^{\bar{b}}d^{\bar{r}}\right) ,\,\,\Lambda
_b^0\left( b^{\bar{b}}d^{\bar{g}}d^{\bar{r}}\right) \rightarrow W^r\Lambda ^{%
\bar{r}}\left( u^gd^{\bar{g}}d^{\bar{r}}\right)  \label{baryondecay}
\end{equation}
where $b^{\bar{g}}\rightarrow W^ru^b$ and $b^{\bar{b}}\rightarrow W^ru^g$ or
vice versa. Let us concentrate upon the above examples of meson decays (\ref
{md1}--\ref{md6}) and baryon decays (\ref{baryondecay}). If we consider
down-type quarks as antiquarks of up-type quarks then in the baryon case $%
\Lambda _b^0$ is composed of three down-type quarks while the $\Lambda ^{%
\bar{r}}$ is composed of one up-type quark and two down-type quarks. Whereas
the baryons are built up of either quarks or antiquarks. By keeping this
view, we cannot consider down-type quarks as antiquarks of up-type quarks.

\section{Conclusions \label{conclusions}}

In this article, the value of the color charges are given. The structure of
the colored, anticolored and color singlet gluons are proposed. Their mass
relations are also given. It is shown that the gauge boson in the
electroweak, QCD and gravity theories are not different but they are linked
to one another. The gauge bosons of lower order theories serve the purpose
for higher order theories. The proof of the claims of Ref. \cite{gilani1}
are given in a systematic way.

Quark fractional charges (i.e. $+\frac 23e$ for up-type quarks and $-\frac
13e$ for up-type quarks) are totally rejected. Only color charges are given
to up-type quarks and anticolor charges to down-type quarks. This is
explained by applying to meson and baryon decays.

{\bf Acknowledgements:} Author thanks to the friends and collegues for
their encourging and/or discourging comments on my previous article \cite
{gilani1}. Author also thanks to those who keep silence.

\section{Figure Captions}

\begin{enumerate}
\item  The plot of cube roots of unity, i.e. the color and anti color
charges.\label{figvalue}

\item  Electroweak gauge bosons $\gamma $, $W^{+}$, $W^{-}$ and $Z^0$.
Photon $\left( \gamma \right) $ gets position at the origin as it is
massless while the others relative to their charge positions. \label
{electroweakbosons}

\item  The gluons are plotted relative to their charge positions. The
massless gluon gets the position at origin. \label{figgluons}

\item  The color points are joined by lines which form triangle $\Delta rgb$
and by joining anticolor points another triangle is formed $\Delta \bar{b}%
\bar{g}\bar{r}$. Both the triangles are equilateral triangles having length
of each side $\sqrt{3}$. \label{cactriangle}

\item  Taking the apexes $r$ and $\bar{r}$ of the triangles $\Delta rgb$ and 
$\Delta \bar{b}\bar{g}\bar{r}$ respectively, which meet at the $z$ on
z-axis. \label{trianglerrbar}
\end{enumerate}

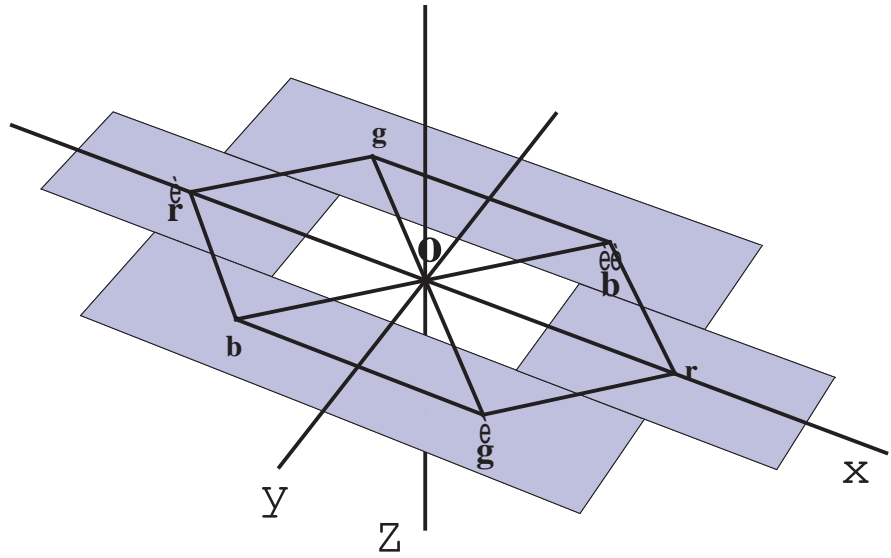
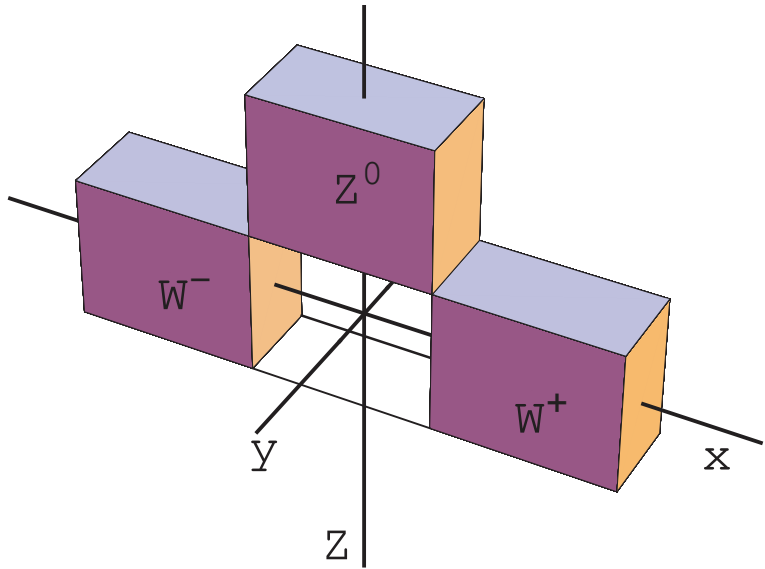
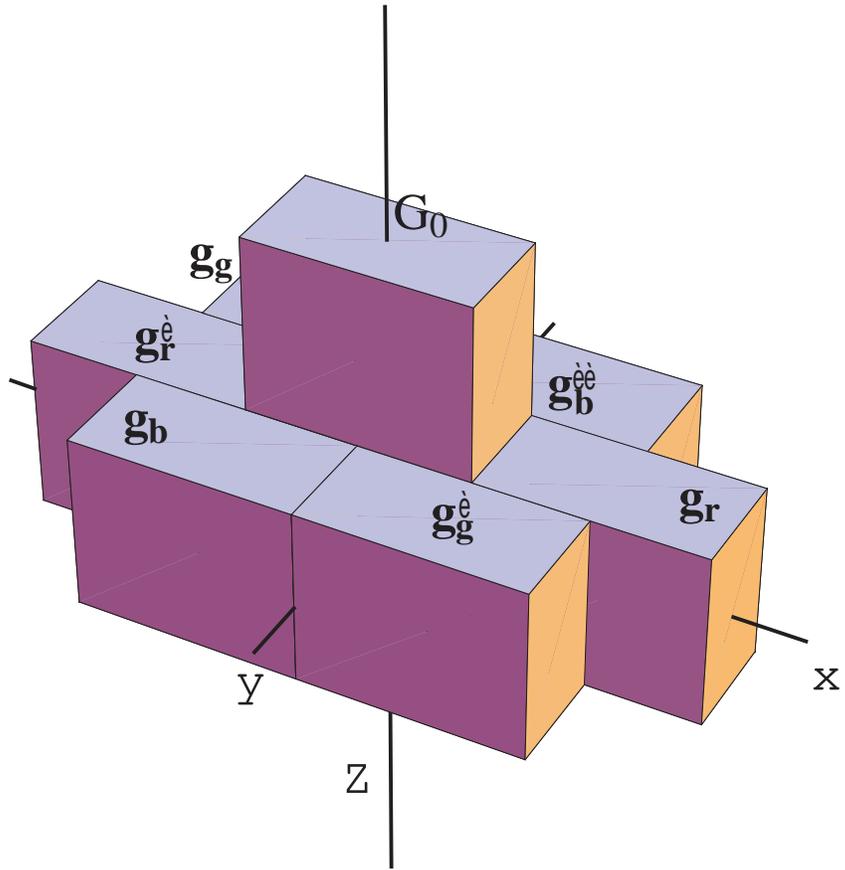
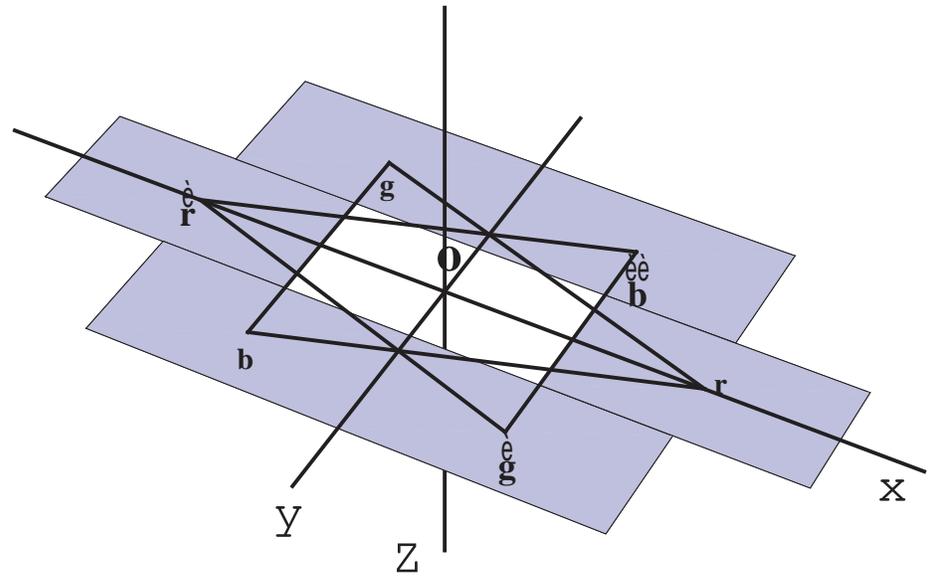
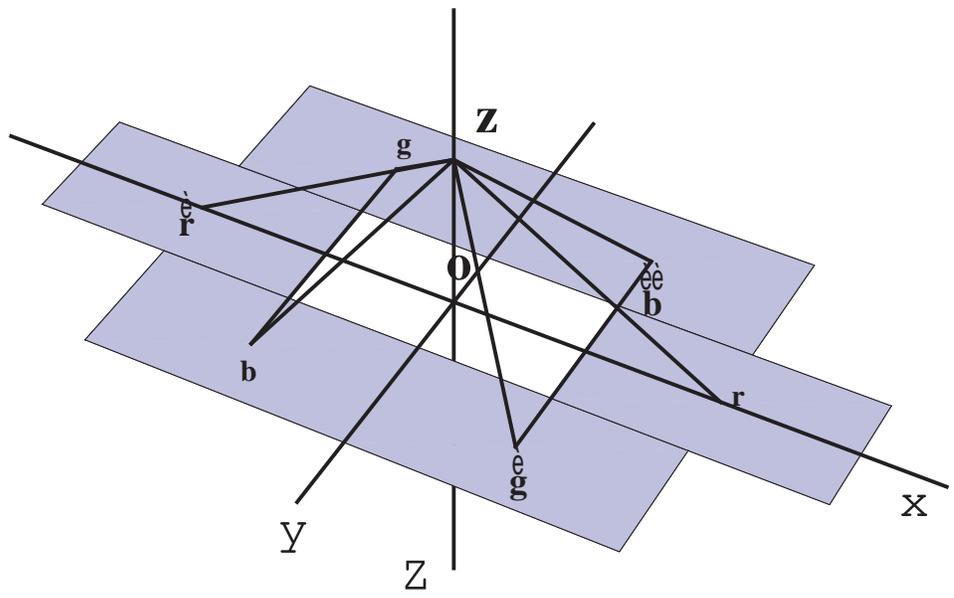
\end{document}